# The Internet of Forgotten Things: European Cybersecurity Regulation and IoT Manufacturer Cessation

Mattis van 't Schip (mattis.vantschip@ru.nl)

Ph.D. Candidate @ Radboud University (iHub)

This is a working draft. I am open to any suggestions and feedback! Please contact me before citing this paper.

## Abstract

Many modern consumer devices rely on network connections and cloud services to perform their core functions. This dependency is especially present in Internet of Things (IoT) devices, which combine hardware and software with network connections (e.g., a 'smart' doorbell with a camera). This paper argues that current European product legislation, which aims to protect consumers of, inter alia, IoT devices, has a blind spot for an increasing problem in the competitive IoT market: manufacturer cessation. Without the manufacturer's cloud servers, many IoT devices cannot perform core functions such as data analysis. If an IoT manufacturer ceases their operations, consumers of the manufacturer's devices are thus often left with a dysfunctional device and, as the paper shows, hardly any legal remedies. This paper therefore investigates three properties that could support legislators in finding a solution for IoT manufacturer cessation: i) pre-emptive measures, aimed at ii) manufacturer-independent iii) collective control. The paper finally shows how these three properties already align with current legislative processes surrounding 'interoperability' and open-source software development.

# 1) Introduction

Internet of Things devices are omnipresent. Organisations and individuals use these devices, which combine traditional hardware (e.g., a watch) with connected functionalities (e.g., WiFi, BlueTooth) for data recording, exchange, and transmission. For instance, individuals might use their 'smart' watch to track and analyse data on their sleep or they might automatically turn on their connected lights when the sun sets.

Many IoT devices rely on their connected functionalities (i.e., the 'Internet' in IoT) to perform core functions.[1] Virtually all IoT devices depend on a connection with the manufacturer's servers to perform data analysis (e.g., indicating sleep health based on data from a smart watch).[2] At the same time, this dependency allows manufacturers to gain user insights and make users dependent on their product ecosystem, as devices can only fully function when connected to the manufacturer's servers.[3]

Next to cloud dependencies, the IoT industry is characterised by a competitive and fast development cycle. Many manufacturers of IoT devices are small or medium-sized enterprises (SMEs).[4] These manufacturers often put devices on the market without adequate consideration for security and privacy, as many they compete for consumers in a busy market.[5] IoT devices can be rather small (e.g., plant sensors) and manufacturers can produce them without large investments.

---

1   Lachlan Urquhart, Tom Lodge and Andy Crabtree, 'Demonstrably Doing Accountability in the Internet of Things' (2019) 27 International Journal of Law and Information Technology 1; Jiahong Chen and others, 'Who Is Responsible for Data Processing in Smart Homes? Reconsidering Joint Controllership and the Household Exemption' (2020) 10 International Data Privacy Law 15.
2   See extensively on the role of the 'cloud' for IoT devices, Milan Milenkovic, *Internet of Things: Concepts and System Design* (Springer International Publishing 2020) 109–152 <http://link.springer.com/10.1007/978-3-030-41346-0> accessed 7 August 2024.
3   W Kuan Hon, Christopher Millard and Jatinder Singh, 'Twenty Legal Considerations for Clouds of Things' (2016) 2016 Queen Mary University of London, School of Law Legal Studies Research Paper.
4   Christopher Boniface, Lachlan Urquhart and Melissa Terras, 'Towards a Right to Repair for the Internet of Things: A Review of Legal and Policy Aspects' (2024) 52 Computer Law & Security Review 105934, 4.
5   Jiahong Chen and Lachlan Urquhart, '"They'Re All about Pushing the Products and Shiny Things Rather than Fundamental Security": Mapping Socio-Technical Challenges in Securing the Smart Home' (2022) 31 Information & Communications Technology Law 99.

These market characteristics have, in certain cases, caused IoT manufacturers to discontinue product development or cease to exist entirely.[6]

If manufacturers discontinue their support, the IoT consumer is often left with little recourse. Consumers of VanMoof, an electric bicycle manufacturer, recently found themselves in this situation. Due to looming insolvency, the cloud-based key required for unlocking consumers' bicycles was no longer accessible, leaving the consumers with doubts whether they could still use their bicycle at all.[7] Consumers had virtually no legal remedies to maintain access to the decryption key required to unlock their bicycles.[8] Without an internet connection to the manufacturer, an Internet of Things device often ends up becoming a regular 'Thing' or entirely obsolete, depending on the product.

The EU legislator aims to significantly reduce the number of unsupported software and hardware devices, mainly by introducing update and support obligations for manufacturers of those devices. These obligations aim to keep products functional and secure for certain periods, so that the manufacturer cannot disregard them, for instance to focus on new product lines. Certain problems, however, remain. Most update and support obligations do not account for the situation in which manufacturers cease to exist entirely, for instance due to bankruptcy. European law does not yet address this problem. Therefore, the main research question of this paper is: How does recent EU legislation regulate IoT manufacturer cessation and which properties from comparative provisions in EU digital policy could support filling any legislative gaps? In this paper, 'manufacturer cessation' refers to manufacturers that cease to exist or cease their business operations entirely, for instance due to bankruptcy.

---

6    Jessica Rich, 'What Happens When the Sun Sets on a Smart Product?' (*Federal Trade Commission*, 13 July 2016) <https://www.ftc.gov/business-guidance/blog/2016/07/what-happens-when-sun-sets-smart-product> accessed 31 January 2024.
7    Ingrid Lunden, 'VanMoof, the e-Bike Startup, Officially Declared Bankrupt in The Netherlands' (*TechCrunch*, 18 July 2023) <https://techcrunch.com/2023/07/18/vanmoof-goes-vanpoof-bankruptcy/> accessed 13 November 2023.
8    A competitor even created an app to help VanMoof customers retrieve their bicycle encryption key from the VanMoof servers shortly before bankruptcy, see Thomas Ricker, 'Cowboy Releases Cheeky App to Keep VanMoof E-Bike Riders on the Road' (*The Verge*, 13 July 2023) <https://www.theverge.com/2023/7/13/23793591/cowboy-vanmoof-key-app-download-ios-android> accessed 31 January 2024.

This paper continues as follows. Section 2) sets out the emerging problem of IoT manufacturer cessation and its risks for consumers. Section 3) focuses on current EU legislation that might address those risks. It argues that current update obligations fall short of providing a framework which fully supports continuous updates and support for IoT devices. The main pitfall of current update obligations is their reliance on the longevity of the manufacturer.[9] For a solution, this paper points towards existing provisions in recent digital policy which require manufacturers to consider their insolvency pre-emptively (Section 4)). Section 5) explores how such pre-emptive measures best fill the legal gap of manufacturer cessation within a i) manufacturer-independent framework that relies on ii) collective control. Such systems already exist and legislators could further use them, as exemplified by the rising attention to interoperability in the Data Act and Digital Markets Act and the reference to open-source code publication in the Cyber Resilience Act. Section 6) concludes.

## 2) Manufacturer Cessation and Its Risks in the Context of the Internet of Things

Manufacturer cessation can take many shapes, depending on why the manufacturer ceases their operations and what type of smart products they produce. In the following, I set out a few examples of IoT manufacturer cessation and how they affected consumers. Next, I categorise the impact of manufacturer cessation on various IoT products and how this impact ranges from a device becoming completely defunct to the device remaining capable of performing its core functions.

### 2.1) The Shapes of Manufacturer Cessation

There are many shapes to manufacturer cessation. First, there are numerous reasons why a manufacturer might cease operations, ranging from an unviable business model to a lack of interest in continuing the business. Second, manufacturers might act differently when deciding to cease operations. Some manufacturers can decide to go through a legal insolvency procedure, while others might instead opt to just cease their operations entirely and liquidate their assets.[10]

---

9   See also Conner Bradley and David Barrera, 'Escaping Vendor Mortality: A New Paradigm for Extending IoT Device Longevity', *New Security Paradigms Workshop* (ACM 2023).

To illustrate these problems, there are several recent examples. In 2023, a Dutch court declared electronic bike manufacturer VanMoof bankrupt.[11] Customers were left in disarray, as they relied on the VanMoof cloud servers to retrieve the encryption key that could, among other functions, unlock their bike.[12] In other words, without access to the VanMoof cloud servers, customers could no longer unlock, and thus use, their ebikes.

More recently, in 2024, Gigaset, a firm that provided smart security cameras, announced that they went into insolvency proceedings. As a result, all devices connected to the Gigaset cloud would no longer be supported and functional, per the Gigaset announcement.[13] Gigaset found a buyer for other parts of the company, but the buyer was not interested in proceeding with the smart home products that Gigaset provided. Customers, therefore, could simply no longer use their devices.

In 2022, consumers of products made by Insteon took matters in their own hands. The company behind Insteon, SmartLights, took their servers offline after financial difficulties, leaving Insteon customers, literally, in the dark as they could no longer access their devices through the app.[14] A group of passionate Insteon users then decided to take over the Insteon business and immediately worked on getting devices back online.

---

10  Cesare Bartolini and others, 'Cloud Providers Viability: How to Address It from an IT and Legal Perspective?' in Jörn Altmann, Gheorghe Cosmin Silaghi and Omer F Rana (eds), *Economics of Grids, Clouds, Systems, and Services*, vol 9512 (Springer International Publishing 2016).
11  Lunden (n 7).
12  Ricker (n 8).
13  The announcement was not provided in English, for the Dutch and German versions see 'Gigaset Smart Home/Care Staakt Zijn Diensten Vanaf 29 Maart 2024' <https://www.gigaset.com/nl_nl/cms/smart-home-overzicht.html> accessed 11 April 2024; Gigaset, 'Gigaset Smart Home/Care Wurde Eingestellt' <https://www.gigaset.com/ch_de/cms/smart-home-uebersicht.html> accessed 16 April 2024.
14  Jennifer Pattison Tuohy, 'Insteon Customers Turned Insteon's Lights Back on' (*The Verge*, 9 June 2022) <https://www.theverge.com/2022/6/9/23161803/insteon-customers-bought-company-restored-service> accessed 31 January 2024.

Another example of manufacturer cessation that attracted much attention is the 'Jibo' robot, which announced its own 'death' when the manufacturer shut down its servers: 'Maybe someday when robots are way more advanced than today, and everyone has them in their homes, you can tell yours that I said hello.'[15] Many consumers became emotionally attached to Jibo over time;[16] they even considered burying it.[17] Additionally, as with Insteon, many Jibo owners attempted to take over Jibo's code base to keep it up and running. In the end, Jibo was also bought by a competitor and kept alive.

These diverse IoT manufacturer cessation examples highlight the dynamics of manufacturer cessation in the Internet of Things sector. All companies suffered from financial difficulties. Consumers were often left without concrete legal remedies, except, perhaps, in the unique Insteon case where they took over the company themselves. The Internet of Things concerns numerous devices, as evident from these examples too: losing access to a smart security camera (Gigaset) or rather expensive devices (VanMoof and Jibo) can be more impactful than losing access to a smart light (Insteon) that, for instance, the consumer can still turn on manually.

## 2.2) Shapes of Product End-of-Support

The examples above show that manufacturers across a range of product types have ceased to exist in previous years. Manufacturer cessation does not impact all products equally. Below, I set out how the impact of manufacturer cessation differs between certain products, ranging from products becoming entirely defunct to products remaining capable of performing their core functions.

---

15  Ashley Carman, 'They Welcomed a Robot into Their Family, Now They're Mourning Its Death' (*The Verge*, 19 June 2019) <https://www.theverge.com/2019/6/19/18682780/jibo-death-server-update-social-robot-mourning> accessed 24 April 2024.
16  Susan Lechelt and others, 'Designing for the End of Life of IoT Objects', *Companion Publication of the 2020 ACM Designing Interactive Systems Conference* (ACM 2020) <https://dl.acm.org/doi/10.1145/3393914.3395918> accessed 24 April 2024.
17  Carman (n 15).

### 2.2.1) Product Becomes Entirely Defunct

Some products rely fully on connecting to the manufacturer's servers for their core functions. These products become entirely non-functional after the manufacturer ceases operations. There are multiple components within an IoT product that can render the device non-functional.

The device can lose access to critical hardware components. This is the case, for instance, when the device requires decryption of data and can only access the decryption key from the manufacturer's servers. As indicated above, customers of the VanMoof eBike could not retrieve their decryption key after VanMoof went into bankruptcy proceedings, thus leaving the consumers unable to unlock their bicycle locks. With the bicycle permanently locked, the bike essentially becomes non-functional for consumers.

More often, there are software functionalities that become defunct after manufacturer cessation, leading to failure for the entire device. Imagine, for instance, that a consumer has a security camera that they can view when they are not home. The application for operating IoT devices is often developed and hosted by the device's manufacturer; data flows from the application to the manufacturer and back.[18] Without the manufacturer's servers, many of the application's functions, including, for instance, a log-in feature for authentication purposes, no longer work.[19] Consumers then lose access to and control over their devices.

### 2.2.2) Products Remain Functional For a While

In the context of IoT manufacturer cessation, products exist in an array from defunct to functional. Smart TVs, for instance, lose some but not all of their key functionalities.

---

18  TJ OConnor, Dylan Jessee and Daniel Campos, 'Through the Spyglass: Towards IoT Companion App Man-in-the-Middle Attacks', *Cyber Security Experimentation and Test Workshop* (ACM 2021).
19  TJ OConnor, William Enck and Bradley Reaves, 'Blinded and Confused: Uncovering Systemic Flaws in Device Telemetry for Smart-Home Internet of Things', *Proceedings of the 12th Conference on Security and Privacy in Wireless and Mobile Networks* (ACM 2019).

Modern smart TVs function quite similar to smartphones as they rely on applications developed by third-parties, such as video streaming apps. The TV can still turn on after manufacturer cessation and consumers can in principle use the TV for any content they desire. For example, they can watch analogue television or even a streaming service.

However, in certain cases, they can no longer use the applications installed on the TV, because using, installing and updating these applications requires a connection with both the TV manufacturer's application store and the application developer's servers. Netflix might, for example, update their interface, which requires certain components from the TV operating system. Consequently, if the TV's operating system is no longer updated by the manufacturer, the Netflix application can no longer function on that TV. In this context, researchers recently found that less than 10% of mobile apps could be downloaded straight from the app store on end-of-support devices (and even fewer actually functioned).[20]

In a similar vein, manufacturers will no longer patch vulnerabilities after they cease their activities. The smart TV therefore becomes vulnerable to cyberattacks, as, over time, numerous vulnerabilities will be discovered in the device's software that attackers could exploit. Manufacturer cessation thus leads to both a loss of functionalities and cybersecurity risks.

Smart TVs and smart phones are examples of devices that thus remain partially functioning, as they, respectively, allow the consumer to watch certain 'analogue' content and call others. In certain cases, applications downloaded from the application store might remain functional even after manufacturer cessation. However, the manufacturer's cessation does lead to loss of functionalities and cybersecurity risks overall. Such devices thus differ from devices that remain fully functional, as discussed below.

---

20  Craig Goodwin and others, 'Quantifying Device Usefulness - How Useful Is an Obsolete Device?' in José Abdelnour Nocera and others (eds), *Human-Computer Interaction – INTERACT 2023*, vol 14145 (Springer Nature Switzerland 2023) 92.

## 2.2.3) Products Remains Capable of Performing Core Functions

There are also certain IoT products that still remain functional after manufacturer cessation.

This is primarily the case when the device can remain functional as a regular Thing. This is often the case for smart lights. Consumers may lose access to the application to remotely operate the lights, but they can still turn them on with a regular wall switch. The wall switch can thus continue to perform the core functions of the lights – offering light – regardless of the manufacturer's cessation.

At the same time, devices such as the smart light also return to a state where they do not differ from traditional, 'dumb' counterparts. Many consumers may therefore still feel a loss of their device, as they bought it precisely because they did not want to manually toggle the lights any more.[21]

Devices may also remain fully functional due to their network architecture. Devices may rely on Wi-Fi-based connections, but could have local network-only options, so that a connection to the manufacturer is not required in the first place. Additionally, devices that do not require a network connection – and thus a connection to the manufacturer – will remain functional after manufacturer cessation through their technical design.[22] Such devices often employ a network protocol other than Wi-Fi, such as BlueTooth, Zigbee, or Z-Wave, which can only communicate with other devices in their proximity. The manufacturer is not within that proximity and thus has no influence over the connection between the consumer's devices. Such a network architecture also reduces security risks, as attackers can also only access the devices when they are in their physical proximity.

---

21 In a recent study, researchers found that 'feeling technologically advanced' and having a 'modern home' were the most beneficial reasons for adopting smart home devices, see Daniel B Shank and others, 'Knowledge, Perceived Benefits, Adoption, and Use of Smart Home Products' (2021) 37 International Journal of Human–Computer Interaction 922, 927.
22 For examples on how these devices remain functional without network connectivity, see Poonam Yadav and others, 'Network Service Dependencies in Commodity Internet-of-Things Devices', *Proceedings of the International Conference on Internet of Things Design and Implementation* (ACM 2019) 208 <https://dl.acm.org/doi/10.1145/3302505.3310082> accessed 4 June 2024.

## 3) EU Product Law in the Context of IoT Manufacturer Cessation

Section 2 highlighted that IoT manufacturer cessation is a diverse problem: manufacturers can cease operations out of nowhere or go bankrupt, often without any consideration for the position of the consumers who bought their devices. Furthermore, the consumer can be left with devices that are entirely defunct or with devices that remain fully functional, depending on the type of product and its technical architecture. In recent years, the EU legislators adopted several pieces of legislation which aim to mitigate cybersecurity, privacy, and safety risks for consumers of IT systems and, in most cases, Internet of Things devices. Below, I examine some legislative solutions to the IoT manufacturer cessation problem and identify their limitations in providing a complete solution.

### 3.1) Update Obligations: Assuming Manufacturer Longevity

Usually, manufacturers publish updates as a way to keep their devices functional and secure. Recent legislation has introduced several update obligations, to ensure that devices remain functional for a longer period. Below, I analyse the update obligations in the Cyber Resilience Act and Sale of Goods Directive in the context of IoT manufacturer cessation.

### 3.1.1) The Cyber Resilience Act

At the beginning of 2024, the European Parliament adopted the 'Cyber Resilience Act'.[23] The Act brings the largest set of cybersecurity responsibilities for manufacturers of software and hardware products—including IoT devices—seen thus far in EU legislation.[24]

---

23 European Parliament legislative resolution of 12 March 2024 on the proposal for a regulation of the European Parliament and of the Council on horizontal cybersecurity requirements for products with digital elements and amending Regulation (EU) 2019/1020 (COM(2022)0454 – C9-0308/2022 – 2022/0272(COD)). The Council still has to adopt this final text, per 21/10/24.
24 Pier Giorgio Chiara, 'The Cyber Resilience Act: The EU Commission's Proposal for a Horizontal Regulation on Cybersecurity for Products with Digital Elements: An Introduction' [2022] International Cybersecurity Law Review.

The scope of the Cyber Resilience Act extends to software and hardware products and their 'remote data processing' capabilities (e.g., a cloud data processing service).[25] The Cyber Resilience Act thus applies to both the IoT device as a hardware/software product and to any cloud services on which it relies.

The Cyber Resilience Act also points to different categories of IoT devices. Next to regular 'products with digital elements', the legislators identified products with additional risks for users. Annex III lists 'important' products, including smart virtual assistants (e.g., Amazon Alexa), smart home products with security functionalities (e.g., Gigaset's cameras), internet connected toys with social or location tracking features and personal wearable products for health monitoring (e.g., smart watches). Additionally, Annex IV lists a few 'critical' products, such as smart meter gateways within smart metering systems, that carry 'significant risks of adverse effects' as they have security functions.[26]

The Cyber Resilience Act defines a certain 'support period' during which manufacturers must ensure that they handle vulnerabilities in their devices effectively, in line with the vulnerability handling requirements in the Annex.[27] The requirements in the Annex include, for example, addressing and remediating vulnerabilities without delay.[28] Pursuant to Article 13(8), the exact period during which manufacturers must offer this support depends on the product, but is usually at least five years.

There are several problems with relying on the Cyber Resilience Act's update requirements to address IoT manufacturer cessation. It is unclear how far this obligation extends if the manufacturer ceases to operate, especially if the manufacturer stops during the support period of the product. It is unlikely that a consumer can effectively rely on market authorities to compel discontinued manufacturers to comply with their support obligations. National law may provide some solutions. For example, provisions may exist that make the organisation's directors liable if they take on obligations which they know they could not comply with because of looming bankruptcy.[29] These provisions differ between Member States and are not harmonised,[30] so they do not (yet) provide a full solution for manufacturer cessation.

---

25  Article 2(1), 3(1), 3(2) & Recital 11-12 CRA.
26  Recital 46 CRA.
27  Art 3(20) & 10(6) CRA.
28  Annex I Part 1(3)(f) CRA.

This lack of remediation points to the assumption of manufacturer longevity on which the support obligations rely. Without the manufacturer, most legal obligations become unviable. Update obligations can therefore only help in remediating problems for IoT devices after manufacturer cessation if such obligations operate independently from the manufacturer (see Section 5)).

For instance, the Cyber Resilience Act states that, after the support period, i.e. the period determined by the manufacturer during which they publish updates for a device, the manufacturer should strive to publish the source code of their product.[31] This option, which only occurs in the Recital and is surprisingly not linked to any actual obligations in the legal text, may allow manufacturer-independent product development, which I discuss further below (Section 5.2)).

The update obligations in the CRA leave a second gap, due to their focus on patching security vulnerabilities. The CRA defines vulnerabilities as 'weakness, susceptibility or flaw of a product […] that can be exploited by a cyber threat.'[32] A cyber threat is an 'event or action that could damage, disrupt, or otherwise adversely impact' a product with digital elements.[33] For certain IoT products, this update obligation misses the mark in the context of manufacturer cessation: if users can still turn on their smart lights manually, they lose certain device functionalities, i.e. control through other network-connected devices, but the devices are not necessarily impacted by security vulnerabilities.

---

29  See for instance the 'Beklamel' jurisdiction in the Netherlands.
30  Carsten Gerner-Beuerle and Edmund-Philipp Schuster, 'The Evolving Structure of Directors' Duties in Europe' (2014) 15 European Business Organization Law Review 191.
31  Recital 62 CRA.
32  Art 3(40) CRA.
33  Art 3(46) CRA & Art 2(8) Regulation 2019/881 (Cybersecurity Act).

The scope of the CRA update obligations therefore fall short in two areas: the type of update and the reliance on manufacturer longevity.

### 3.1.2) Sale of Goods Directive

The Sale of Goods Directive (SGD) is in force since 2019.[34] The SGD ensures consumer protection concerning sale contracts between 'sellers' and 'consumers'. Sellers are natural or legal persons who act on behalf of their business or trade, while consumers are natural persons who are not acting for purposes of their business or trade.[35] The SGD concerns contracts for the sale of 'goods', which also include Internet of Things devices, as 'goods with digital elements'.[36]

The SGD contains two categories of conformity requirements: subjective and objective. Article 6 lists several subjective requirements, such as i) the possession of certain functionalities and compatibilities, as required by the sales contract, ii) conformity to the consumer's purposes and iii) the supply of updates as stipulated by the contract. Subjective requirements thus depend on the sales contract between the seller and consumer.

Article 7 lists objective requirements, which apply irrespective of what is stipulated in the sales contract. The seller must comply with both the subjective and objective requirements. Update requirements are also part of these objective requirements.[37] The subjective update requirements stem from the sales contract, while the objective update requirements ensure that goods with digital elements remain in conformity, i.e. comply with these requirements and certain installation requirements,[38] for at least a certain period of time.[39] This period is determined, inter alia, by the reasonable expectations of the consumer.[40]

---

34  Directive (EU) 2019/771 of the European Parliament and of the Council of 20 May 2019 on certain aspects concerning contracts for the sale of goods, amending Regulation (EU) 2017/2394 and Directive 2009/22/EC, and repealing Directive 1999/44/EC [2019] OJ L136/28.
35  Art 2(2) and 2(3) SGD.
36  Art 2(5)(b) SGD.
37  Art 7(3) SGD.

A key difference with the Cyber Resilience Act is that the SGD's update requirements refer to the conformity of the product with the sales contract.[41] The updates must thus ensure that the device keeps providing, for instance, the functionality, compatibility, and interoperability as agreed in the sales contract.[42] The Directive also explicitly mentions security updates.[43] The CRA, in contrast, aims exclusively at mitigating security vulnerabilities. For manufacturer cessation, the SGD addresses another side of the manufacturer cessation problem: the degrading functionalities of the IoT device after the manufacturer ceases operations. Here, the SGD thus complements the CRA regarding the types of updates.

The SGD requires the seller to provide updates, instead of the manufacturer in the Cyber Resilience Act. If the seller is not the manufacturer, it lacks the necessary expertise. The seller then depends on co-operation with the manufacturer.[44] Consequently, the update requirement in the SGD relies on the same assumption of manufacturer longevity as the Cyber Resilience Act. When the manufacturer ceases to exist, holding the seller responsible for updates is no longer a viable option. The consumer can hold the seller liable for the lack of updates or claim other remedies,[45] but that does not change the fact that the device is no longer supported. The consumer is thus left without updates regardless of the responsible party. The update requirements in the SGD and CRA thus highlight that a full solution to the problem of IoT manufacturer cessation cannot rely on the assumption of manufacturer longevity.

---

38  Art 5 SGD.
39  Art 7(3) SGD.
40  Art 7(3)(a) and (b) SGD. On consumer expectations, see Simon Geiregat, 'What Digital Content Consumers (Should) Want' in Anna K Bernzen, Karina Grisse and Katharina Kaesling (eds), *Immaterialgüter und Medien im Binnenmarkt* (Nomos Verlagsgesellschaft mbH & Co KG 2022).
41  Art 5 SGD.
42  Art 6(a) SGD.
43  Art 7(3) SGD.
44  See Piia Kalamees, 'Goods With Digital Elements And The Seller's Updating Obligation' (2021) 2 JIPITEC 131, 134; Hans Schulte-Nölke, 'Digital Obligations of Sellers of Smart Devices under the Sale of Goods Directive 771/2019' in Sebastian Lohsse, Reiner Schulze and Dirk Staudenmayer (eds), *Smart Products* (Nomos 2022).
45  Art 10 & 13 SGD.

## 4) Moving Away from Manufacturer Longevity Based on Existing Digital Policy

Update obligations assume that the manufacturer remains capable of supporting their products. In the case of manufacturer cessation, the manufacturer loses that capability. If the manufacturer cannot act after cessation (*ex post*), pre-emptive measures (*ex ante*) may offer a viable alternative. At the same time, those pre-emptive measures only work if the manufacturer, after cessation, no longer carries responsibility for the device. Therefore, a solution to manufacturer cessation also requires *manufacturer-independency*. Finally, there should be room for collective control, as continuing to rely on a single responsible party retains the risks of manufacturer dependencies, i.e. cessation of the responsible party. These properties are also present in several pieces of recent EU digital policy.

### 4.1) Cyber Resilience Act

The Cyber Resilience Act requires manufacturers to inform relevant market surveillance authorities and, by any means available and to the extent possible, the consumers of their devices when they cease their operations.[46] This requirement only applies when the manufacturer, because it ceases to exist, can no longer comply with the Cyber Resilience Act. Importers and distributors of the manufacturer's devices have the same obligation when they receive information that the manufacturer ceases or has ceased to exist.[47] The Cyber Resilience Act thus contains a *pre-emptive measure* in the form of a notification mechanism. This notification mechanism, however, can only serve to reduce the risks of manufacturer cessation for consumers, as highlighted in Section 2.2), since the notification allows consumers to prepare and take measures (e.g., stop using their device).

---

46  Art 13(23) CRA.
47  Art 19(8) CRA.

The Recital of the Cyber Resilience Act suggests that manufacturers publish the source code of their product publicly after the end of the support period, or transfer it to another company for vulnerability handling.[48] Over the past decades of digital product development, some consumers with software development experience have advocated for publication of code, as it allows such experienced developers to continue product development even if the manufacturer decides to end their support.[49] As such, open-source code publication could serve a *manufacturer-independent* solution within a development framework of *collective control*: many open-source developers together build new software for the device. There is also some potential in transferring the source code to another company, as the Recital suggests, which I discuss further in Section 5.3). The Cyber Resilience Act thus also indicates, options for device continuity after manufacturer cessation.

### 4.2) Digital Operational Resilience Act

The Digital Operational Resilience Act (DORA), the EU cybersecurity law for the financial sector, includes pre-emptive obligations for insolvency.[50] Article 30 notes several contractual provisions that are mandatory when a financial entity (e.g., a bank) employs the services of a third-party ICT provider. The financial entity must include provisions which ensure that personal data processed by the third-party ICT provider is returned 'in the event of the insolvency, resolution or discontinuation of the business operations' of the provider.[51] The financial entity and the third-party ICT provider must thus together consider the possible insolvency of the ICT provider when signing the contract.

---

48  Recital 62 CRA.
49  Joe Rickerby, 'Resilient Products — How Connected Devices Can Live On.' (*Nord Projects*, 17 May 2019) <https://medium.com/nord-projects/resilient-products-how-connected-devices-can-live-on-b9c0bb9534e6> accessed 24 April 2024.
50  Regulation (EU) 2022/2554 of the European Parliament and of the Council of 14 December 2022 on digital operational resilience for the financial sector and amending Regulations (EC) No 1060/2009, (EU) No 648/2012, (EU) No 600/2014, (EU) No 909/2014 and (EU) 2016/1011 [2022] OJ L333/1.

As in the Cyber Resilience Act, the DORA thus requires certain *pre-emptive* considerations for insolvency of a party that provides ICT services. The CRA only imposes responsibilities on the manufacturer when they cease operations. The DORA insolvency rules are slightly stronger as parties must also consider insolvency upon signing the agreement, when insolvency is not a looming problem yet.

The return of data is a measure that could hold value for manufacturer cessation too. For example, the manufacturer could, at the very least, ensure that (personal) data processed by a device is sent back to the consumer when they cease their operations. This data holds value for device continuity, or at least the continuity of the service (e.g., tracking sleep data over several years). Therefore, data return also allows for a *manufacturer-independent* solution based on data.

### 4.3) Data Governance Act

The Data Governance Act, which, inter alia, regulates 'data intermediation services', also notes that these services must ensure continuity of their services in case of their insolvency.[52] There is no clarity in the DGA about how the services must ensure this continuity, leaving options open to the service itself. It is imaginable that services might open a portal through which users can access their data for a limited period, or that they aim to co-operate with other, similar services that could continue to support their infrastructure. These actions also depend on the national insolvency procedures, because, in certain Member States, the service owner cedes authority to a liquidator or insolvency administrator once they enter into insolvency proceedings.[53] The DGA thus brings a pre-emptive measures that aims at continuation, where the CRA and the DORA consider insolvency the end of the service. The DGA therefore offers the most compelling type of *pre-emptive measure* in the context of manufacturer cessation: a measure that appreciates the lifespan of manufacturers.

---

51 Art 30(2)(d) DORA.
52 Art 12(h) Data Governance Act.
53 As also identified in Lukas Von Ditfurth and Gregor Lienemann, 'The Data Governance Act: – Promoting or Restricting Data Intermediaries?' (2022) 23 Competition and Regulation in Network Industries 270, 290.

## 4.4) Interoperability policies

In several pieces of legislation, the EU legislators emphasise the need for more interoperability in the current data economy. Interoperability, generally, refers to the capacity of systems to co-operate. Indeed, when systems can co-operate, it is easier to create a manufacturer-independent solution after their cessation. The consumer could, for instance, integrate the device into another system that could exchange data with the device. This type of integration also creates a framework of collective control where multiple devices, systems, and infrastructures co-operate within the framework. I discuss this design more extensively below in Section 5.1).

Interoperability occurs in a myriad of EU policies.[54] Below, I address the pieces most relevant for the (domestic) IoT context and the position of the consumer in case of manufacturer cessation.

### 4.4.1) General Data Protection Regulation

The General Data Protection Regulation is the key EU policy for data protection. The GDPR prescribes specific rights to individuals whose data is 'processed', i.e. the consumer of IoT products. One of those rights is the right to data portability.[55]

The GDPR introduces the right to data portability to strengthen the position of the data subject vis-à-vis the data controller.[56] The right to data portability means that data subjects can receive the data concerning them from the entity holding that data, the 'controller' under the GDPR, in a 'commonly used and machine-readable format'.[57] This provision should, according to the Recital accompanying it,[58] encourage data controllers to adopt interoperable data formats.[59]

---

54 For instance the 'Interoperability Regulations' in the field of digital migration policy: Regulation (EU) 2019/817 of the European Parliament and of the Council of 20 May 2019 on establishing a framework for interoperability between EU information systems in the field of borders and visa (Border Interoperability Regulation) [2019] OJ L135/27.
55 Art 20 GDPR.
56 Recital 68 GDPR.

Data portability or interoperability holds potential for strengthening the position of the IoT consumer.[60] With data portability, the consumer can more easily switch between services, as they can transfer their data from one manufacturer to another, or from one ecosystem to another. This easier exchange also manifests, as highlighted earlier, the manufacturer-independent characteristics of interoperability. Importantly, however, the right to data portability under the GDPR only concerns the transfer of personal data that were 'provided' by the data subject.[61] Algorithmically inferred statistics, for instance, are not part of the right to data portability.[62]

### 4.4.2) Data Act

The Data Act addresses several elements of data availability and requires the interoperability of data between different entities.[63] The DA's rules on interoperability focus on the interoperability of 'data processing services', especially if they cover the same service type.[64] The DA defines data processing services as services which enable 'ubiquitous and on-demand network access to a shared pool of […] computing resources', i.e. cloud services.[65] The DA introduces interoperability as a *manufacturer-*

---

57  Art 20(1) GDPR.
58  Recital 68 GDPR.
59  See more extensively, Paul De Hert and others, 'The Right to Data Portability in the GDPR: Towards User-Centric Interoperability of Digital Services' (2018) 34 Computer Law & Security Review 193.
60  Lachlan Urquhart, Neelima Sailaja and Derek McAuley, 'Realising the Right to Data Portability for the Domestic Internet of Things' (2018) 22 Personal and Ubiquitous Computing 317.
61  See on the 'provided by' condition the guidelines from the Article 29 Working Party: Article 29 Data Protection Working Party, 'Guidelines on data portability' 9-11.
62  Urquhart, Sailaja and McAuley (n 60) 326.
63  Regulation (EU) 2023/2854 of the European Parliament and of the Council of 13 December 2023 on harmonised rules on fair access to and use of data and amending Regulation (EU) 2017/2394 and Directive (EU) 2020/1828 (Data Act).
64  Art 35 DA.
65  Art 2(8) DA.

*independent* tool, which can 'overcome vendor lock-in',[66] because interoperability allows two or more systems to 'exchange and use data in order to perform their functions.'[67]

### 4.4.3) Data Governance Act

The Data Governance Act is closely aligned with the Data Act and therefore also provides certain interoperability obligations.

Mainly, the data intermediation services should integrate interoperability standards into their operations. Next to their obligations for continuity after insolvency, Article 12 lists that the services must enhance interoperability across different sectors, for instance by converting data into certain formats.[68] They must also ensure interoperability between different data intermediation services.[69] It is imaginable that such interoperability between services also helps the continuity of the data intermediation service after insolvency, as the other service can more easily work with the data processed by the insolvent service.

### 4.4.4) Digital Markets Act

The Digital Markets Act applies since May 2023.[70] The DMA requires 'gatekeepers', the largest global online platforms (e.g., Meta, Google),[71] to incorporate interoperability into their systems.[72]

In this context, interoperability should serve consumers of, for example, messaging apps, so that they can use the messaging service that they most prefer, instead of being locked in to specific vendors (e.g., WhatsApp by Meta). Interoperability in such terms is often termed as 'federation':

---

66   Recital 90 DA.
67   Art 2(40) DA.
68   Art 12(d) DGA.
69   Art 12(i) DGA.
70   Regulation (EU) 2022/1925 of the European Parliament and of the Council of 14 September 2022 on contestable and fair markets in the digital sector and amending Directives (EU) 2019/1937 and (EU) 2020/1828 (Digital Markets Act) [2022] OJ L265/1 (DMA). For application date, see Article 54 DMA.
71   European Commission, 'Commission Designates Six Gatekeepers under the Digital Markets Act - European Commission' <https://digital-markets-act.ec.europa.eu/commission-designates-six-gatekeepers-under-digital-markets-act-2023-09-06_en> accessed 9 February 2024.
72   Art 6(7) DMA.

people can choose their own digital service provider, without having to follow their peers or feel pressured by the network effects of dominant platforms.[73] Instead, any digital service provider or environment they choose offers the capacity to co-operate with other service providers (e.g., how e-mail providers co-operate: Gmail and Outlook users can exchange messages). The DMA thus introduces interoperability as a component of its aims to curtail the market power of gatekeepers.[74]

## 4.4.5) Interoperability as a legal solution to manufacturer cessation

Interoperability may hold potential for IoT devices after manufacturer cessation, depending on legal incentives. Many parties must be willing to co-operate and ensure that the product is capable of communicating with other devices and shared platforms. If manufacturers do not pre-emptively ensure such interoperability, the responsibility for ensuring interoperability lies on enthusiastic users and other developers. Manufacturer interest in interoperability is often low, as they would rather keep consumers inside their product ecosystem, as illustrated by the reasoning behind interoperability in the legislation above. However, when designed as pre-emptive measures, interoperability offers an apt fallback in case of IoT manufacturer cessation.

Interoperability after manufacturer cessation is therefore, in most cases, only a practical solution if implemented as a *pre-emptive legal obligation* for manufacturers. Following this obligation, interoperability then provides a *manufacturer-independent* tool for *collective control* of the IoT device.

---

73 Ian Brown, 'Interoperability as a Tool for Competition Regulation' (LawArXiv 2020) preprint 8 <https://osf.io/fbvxd> accessed 6 February 2024.
74 Recital 57 DMA.

|  | Pre-emptive measures | Manufacturer-independency | Collective control |
|---|---|---|---|
| Cyber Resilience Act | Notification mechanism when manufacturer ceases operations | Open-source code publication when ceasing product support | |
| Digital Operational Resilience Act | Contract between financial entity and third-party ICT provider must consider insolvency of the latter | Return of (personal) data after insolvency | |
| Data Governance Act | 1. Data intermediation service must ensure continuity of service upon insolvency 2. Interoperability measures | Interoperability measures | |
| General Data Protection Regulation | Interoperability measures | | |
| Data Act | | | |
| Digital Markets Act | | | |

*Table 1: Recent EU digital policy initiatives that might help against manufacturer cessation problems*

## 5) Designing IoT in the face of manufacturer cessation

Existing update requirements that aim to ensure device maintenance rely strongly on the longevity of the manufacturer, as shown by the Cyber Resilience Act and the Sale of Goods Directive. Such approaches do not offer a solution for consumers who are left with an Internet of Things device created by a non-existent manufacturer. Below, I further analyse how a legislative solution can circumvent manufacturer longevity, while employing pre-emptive measures.

Chapter 4) highlights that the insolvency of a service provider or manufacturer in the digital context is in fact a concern of the European legislators. The provisions show that the European legislators aim to alleviate some of the problems caused by insolvency by requiring pre-emptive measures, ranging from information obligations upon *looming* insolvency (Cyber Resilience Act), to measures upon signing a contract for *possible future* insolvency (Digital Operational Resilience Act), and concerns for the continuity of the service *after* insolvency (Data Governance Act).

Manufacturer independency is the focal point of such solutions. The manufacturer ceasing to exist is the exact cause of the problems for the IoT consumer. Therefore, the manufacturer should set in place pre-emptive measures, but should not be responsible for the continuation of the device after their cessation.[75]

Collective control refers to the capacity of a multitude of actors to shape the continuity of the device. The single dependency on the manufacturer has shown its pitfalls when the manufacturer ceases to exist and therefore a solution should rely on a more decentralised support model, without a single point of failure.[76]

Below, I show examples of IoT design which incorporates several of these components, based on the policies analysed in Chapter 4). These examples illustrate in which direction the legislator could move in designing a solution for IoT manufacturer cessation problems. Given the significant legislative attention to interoperability, I first consider this option.

---

75 Bradley and Barrera (n 9); Ghazaleh Shirvani and Saeid Ghasemshirazi, 'Towards Sustainable IoT: Challenges, Solutions, and Future Directions for Device Longevity' (arXiv, 2024) <https://arxiv.org/abs/2405.16421> accessed 25 June 2024.
76 Bradley and Barrera (n 9).

## 5.1) Interoperability

Interoperability relates to the capacity of systems to work together, regardless of their manufacturers or vendors.[77] As shown above, the EU legislator often utilises interoperability as a tool against platform power, for instance in the recently adopted Data Act (DA) and Digital Markets Act (DMA). Here, I address how interoperability works specifically for the Internet of Things.

### 5.1.1) A Taxonomy of Interoperability for the Internet of Things

The Data Act and Digital Markets Act highlight that interoperability is a dynamic concept. Interoperability exists across various dimensions. These three dimensions do not operate independently, but highlight which choices come to the fore when developers or manufacturers aim to establish interoperability.

**Manufacturer interoperability**

Interoperability can be horizontal or vertical. First, manufacturers can allow interoperability of the devices that they produce (e.g., Apple devices having access to Apple Messenger). This mode of interoperability is called 'vertical' interoperability. Second, manufacturers can allow interoperability between their systems, devices, and platforms and those of other manufacturers, a mode called 'horizontal' interoperability. In the case of Apple Messenger, horizontal interoperability would mean that users could use devices (e.g., Android phone) or services (e.g., Signal) from any manufacturer to communicate through Apple Messenger. The same goes for IoT: horizontal interoperability means that a smart voice assistant from Google can turn on a smart light from Philips.

---

77 Brown (n 73); This is a general notion of interoperability. For further technical and non-technical definitions, see the discussion in Jörg Hoffmann and Begoña Gonzalez Otero, 'Demystifying The Role Of Data Interoperability In The Access And Sharing Debate' (2021) 11 JIPITEC 252, 255–257.

**Data exchange interoperability**

If manufactures aim to make their systems interoperable, they must consider how the interoperable systems exchange data. Two interconnected types of interoperability exist in the context of the technical design of interoperability: syntactic and semantic interoperability.

Syntactic interoperability generally refers to data exchange interoperability.[78] To achieve syntactic interoperability in the Internet of Things, it is imperative that devices send and exchange data in commonly used data formats, such as XML or JSON.[79] The intention of such formats is to display data in a machine-readable format, in contrast with common data formats that aim to display data in a legible way for users (e.g., a PDF file). Through machine-readable formats, syntactic interoperability then allows data conversion between different units, such as temperatures recorded in Celsius by one device and in Fahrenheit by another.

Semantic interoperability refers to the capacity of systems to work with a common, shared vocabulary. Even when systems are syntactically interoperable, their definitions within the shared data sets can vary significantly. For instance, the manufacturer of one device might record temperature as 'heat', while the other records temperature as 'room_temperature'. Syntactic and semantic interoperability thus collaborate: through common machine-readable formats and vocabularies, devices and systems can more easily interoperate.

**Methodology of interoperability**

Interoperability can exist in a manufacturer's systems, processes, and devices. For Internet of Things devices, there are often two prevalent options: device interoperability and platform interoperability. Recently, researchers also suggested a third option in the context of device continuity: update interoperability.[80]

---

78 Mahda Noura, Mohammed Atiquzzaman and Martin Gaedke, 'Interoperability in Internet of Things: Taxonomies and Open Challenges' (2019) 24 Mobile Networks and Applications 796, 799.
79 ibid.
80 Bradley and Barrera (n 9).

Device interoperability depends on the actual device itself: the device can co-operate with other devices, for instance based on the use of similar wireless communication protocols. As mentioned earlier, Internet of Things devices exchange and transmit data through several protocols, such as Wi-Fi, Zigbee, and Z-Wave.[81] For device interoperability, it is thus necessary that, among such diverse protocols, devices can still communicate and exchange data.[82] Device interoperability thus depends on device-to-device interoperability.

Interoperability can depend on different levels of a device's software and hardware stack. For instance, as mentioned above, interoperability can exist on the level of communication protocols, such as Zigbee or BlueTooth. Simultaneously, the communication cannot be interoperable at all, but the device may use a USB-C port for charging, which is, in a sense, interoperable with other USB-C devices. Here, there are thus also differences between hardware and software interoperability, with the former being much easier than the latter, as evident from the transition to USB-C charging in the EU.[83]

Platform interoperability is a method for interoperability that does not depend on the device, but on an integrated platform through which multiple systems can interoperate. Smart home platforms such as Home Assistant and openHAB are examples of interoperable Internet of Things platforms.[84] These platforms allow users to integrate numerous devices from various manufacturers to exchange data. Developers ensure that the data from device A is syntactically and semantically capable of interacting with device B. For this data conversion, the platform may rely on Application Programming Interfaces (APIs), a software interface that allows easy access to a device's data. In addition, the platform allows communication between devices that communicate through differing wireless protocols. For example, the platform ensures that a *Zigbee* motion sensor can ensure that a *Wi-Fi-based* light turns on.

---

81 Eyhab Al-Masri and others, 'Investigating Messaging Protocols for the Internet of Things (IoT)' (2020) 8 IEEE Access 94880.
82 Noura, Atiquzzaman and Gaedke (n 78) 798.
83 Directive (EU) 2022/2380 of the European Parliament and of the Council of 23 November 2022 amending Directive 2014/53/EU on the harmonisation of the laws of the Member States relating to the making available on the market of radio equipment [2022] OJ L315/30.
84 For an overview of these and similar open-source smart home platforms, see Brian Setz and others, 'A Comparison of Open-Source Home Automation Systems' (2021) 9 IEEE Access 167332.

Through platform interoperability, the maintenance of IoT devices no longer relies on a single manufacturer. On an interoperable platform, data analysis happens through the platform. Device updates are necessary only for keeping the data flow to the platform functional. As such, platform interoperability allows for a solution independent from the manufacturer, as the platform's maintainers aim to ensure that they can read the data from the device in the manner necessary to keep it interoperable with the platform. For instance, a smart watch should send the correct sleep data, but the platform works with the data to indicate sleep level and health. In essence, the interoperable platform could thus replace the manufacturer's servers in terms of data analysis.

In the case of IoT manufacturer cessation, Bradley and Barrera suggest a different type of interoperability: update pipeline interoperability.[85] In their framework, IoT software update pipelines are interoperable: devices can pull their updates from a range of repositories. Bradley and Barrera suggest that manufacturers pre-emptively ensure that their device is capable of detecting when they cease their support: 'instead of an unsupported device running progressively more outdated software, it can decide when to transition and switch to actively maintained software to avoid suffering from software degradation.'[86] The device thus interoperates with multiple update pipelines, ensuring manufacturer-independent update continuity.[87]

| Manufacturer | Data exchange | Methodology |
| --- | --- | --- |
|  |  |  |

---

85 Bradley and Barrera (n 9) Bradley and Barrera do not define their framework within the context of 'interoperability' themselves, but their framework has a lot of interoperable characteristics.
86 ibid 11.
87 ibid 12.

| Vertical | Syntactical | Platform |
|---|---|---|
| Horizontal | Semantic | Device (software/hardware) |
| | | Update |

*Table 2: Dimensions of interoperability*[88]

## 5.1.2) Designing Interoperability as a Solution to Manufacturer Cessation

In the face of manufacturer cessation risks, interoperability requires certain design considerations, per the taxonomy above.

First, preferably, the device can communicate with an interoperable platform, with device-to-device interoperability as a secondary option. An interoperable platform allows for the connection between heterogeneous devices and systems, thus simply covering a broader, horizontal scope than device-to-device interoperability.

Second, and more complex, interoperability should help ensure device longevity through updates. There are two levels of updates required for the device: functionalities and security. In terms of functionality with the platform, the platform maintainers themselves could ensure that they keep the device functional with the platform. This is most likely only possible where the device's code base is open-source, as maintainers cannot ensure functionality without knowing how the device records and exchanges data. The maintainers should primarily address issues of data exchange, for example because of API changes, so that they keep access to the device's data for analysis on the platform.

Security updates, however, are most likely fully outside the reach of the platform maintainers, as they may depend on the device's operating system. Here, the suggestions described above for update pipeline interoperability may come into play. A device can remain secure by pulling operating system updates, which include security updates, from sources other than the original manufacturer.[89] In an ideal situation, the manufacturer would pre-emptively support these capabilities for functionality and security updates, i.e. by setting up data conversion and update pipeline interoperability.

---

88  These taxonomies are derived from Noura, Atiquzzaman and Gaedke (n 78).

Given the caveats of device updates, interoperability cannot (yet) offer a fully independent solution to IoT manufacturer cessation. A combination with other solutions could receive further attention, especially, as mentioned above, with open-source code publication.

## 5.2) Open-Source Publication and Development

Source code publication holds potential in the context of IoT manufacturer cessation. For developers, the source code of a device reveals its inner workings. For instance, the source code tells developers how a device captures data, how it performs certain data analysis on the cloud, and what type of security measures the device uses. With the source code in hand, developers can therefore continue developing the device, for instance by adding new features or removing cloud dependencies and creating local alternatives.[90] Source code publication thus allows for manufacturer-independent collective control, as all interested developers (if they exist) can download the code and continue development according to their own needs and interests.

However, there are several barriers standing in the way of open source publication.[91] As the CRA only includes open source publication in the Recital, manufacturers do not have a legal *obligation* to release their source code online. For an IoT manufacturer, the device's source code is often the foundational intellectual asset for the entire company. Manufacturers are thus generally hesitant to release it freely. It is imaginable that the manufacturer wants to keep the code secret so that they can, for instance, sell it during insolvency proceedings.

---

89  Bradley and Barrera (n 9) 12.
90  Rickerby (n 49).
91  Bradley and Barrera (n 9) 4–5.

Additionally, source code release brings intellectual property considerations. The code could, for instance, partially rely on other software which is protected by intellectual property rights (e.g., copyright).[92] These intellectual property rights may stifle attempts to publish the source code publicly.

In face of these challenges, some researchers have suggested that software escrows might offer a solution for software business continuity.[93] Software escrow exists in different forms. Some software escrows exist in the form of 'source code escrow', where the escrow agency releases the source code of the software after the developer ceases to exist. This form is very rarely seen, given the complex demands of source code publication.[94]

However, given the challenges posed by IoT manufacturer cessation and the interest in source code publication in the Cyber Resilience Act, source code escrow could be re-considered. Source code escrow requires pre-emptive action: the manufacturer works with an escrow agency to ensure the release of their source code based on certain conditions, such as bankruptcy. Within this agreement, they can also pre-emptively ensure the correct transfer of intellectual property rights with any rights holders, or ensure that certain parts of the code are adjusted so that intellectual property issues are prevented.

Software escrow can also exist as a form of insurance.[95] For instance, for the software-as-a-service industry, escrows mainly ensure that the developer's hosting costs are paid for a certain period after they cease their activities.[96] In this role, software escrow works more as an insurance fund for business continuity, depending on the contracts agreed by the manufacturer.

---

92  P McCoy Smith, 'Copyright, Contract, and Licensing in Open Source' in Amanda Brock (ed), *Open Source Law, Policy and Practice* (2nd edn, Oxford University Press 2022).
93  Tommy Van De Zande and Slinger Jansen, 'Business Continuity Solutions for SaaS Customers' in Björn Regnell, Inge Van De Weerd and Olga De Troyer (eds), *Software Business*, vol 80 (Springer Berlin Heidelberg 2011); Bartolini and others (n 10).
94  See also more concerns in Jonathan L Mezrich, 'Source Code Escrow: An Exercise in Futility?' (2001) 5 Marquette Intellectual Property Law Review 117.
95  Bartolini and others (n 10) 291–293.

An insurance fund also supports the need for a hosting solution for the open-source software after code publication. There are several possibilities here: first, developers not only continue development, but also host a server that performs similar functions to the insolvent manufacturer's servers; second, the escrow insurance fund ensures that hosting costs are paid, but this only covers hosting for a certain time, thus eventually standing in the way of device continuity; third, consumers can connect to a local server, such as the interoperable smart home platforms mentioned earlier, and get the open-source software from there or by transferring it from another system to their IoT device (e.g., their computer with a cable).

Source code publication offers a method that fits well within the framework suggested in this paper: it is a measure that manufacturers can i) pre-emptively set in place, for instance through source code escrow. After cessation, the source code allows other developers to sustain the device's software or create new software: a ii) manufacturer-independent mode of iii) collective software development. The legislators can move forward from the Cyber Resilience Act's nudge towards source code publication by overcoming the challenges posed above, particularly through concrete legal obligations for manufacturers. Furthermore, the legislators could explore the potential of combining interoperability (as an alternative data platform) and open-source code publication (for device longevity through updates).

## 5.3) Suggestions for Future Research

Above, I proposed a framework for future legislative solutions to the IoT manufacturer cessation problem. Given recent regulatory efforts, it seems that interoperability and open-source code publication, ideally in conjunction, hold most potential as a solution to manufacturer cessation. However, these solutions have certain shortcomings, as IoT manufacturer cessation is a complex multi-faceted issue. Therefore, I propose several other, more novel solutions that future research could utilize as groundwork for finding a legal pathway to tackling IoT manufacturer cessation problems.

---

96  Van De Zande and Jansen (n 92) 23.

First, market supervisors could require competitors to continue the business operations of their insolvent peers in a caretaker role. This could be a pre-emptive measure – if your product is brought to the European market, you must have a competitor able to take over your product – or a mandatory measure imposed by supervisory authorities on competitors. In the European banking sector, such forced take-overs are part of the powers of the banking supervisors.[97]

At the same time, the banking sector differs from the IoT sector. The number of banks is limited, while the IoT sector consists of numerous manufacturers from diverse countries that bring very diverse products to the European market. Banks cannot simply enter the market, they have requirements for capital and require certain permits. IoT manufacturers are often just one part of a larger technology manufacturer. The 'caretaker' scheme could therefore only work if the caretaker role is placed fully externally, instead of with another sister organisation that might suffer the same fate as the original manufacturer. Here, the escrow option – an external party holding the source code – most likely works better than a mandated transfer to another company. In this line, it is also fruitful to further explore the transfer of source code to an external 'caretaker', as suggested by the Recitals of the Cyber Resilience Act.

Second, researchers and legislators could consider the option of a mandated local alternative for Internet of Things devices. This mandate would mean that devices must also have the capability to function without a connection to the manufacturer, through local-only connections. In the smart home context, there are already plenty of examples of such devices. With a local-only mode, the cessation of a manufacturer poses less of a threat to consumer devices, as the consumer can switch to the local-only alternative to keep the device functional. At the same time, the consumer still requires a solution for the lack of updates stemming from the manufacturer's cessation.

---

97 See the 'sale of business tool' in Article 38 & 39 of Directive 2014/59/EU (Bank Recovery and Resolution Directive).

Third, within the realm of open-source code publication, there are multiple cases for future research. Primarily, future research could focus on how to ensure that not only updates, but also hosting is ensured after open-source code publication, based on some of the examples mentioned above. In addition, given the barriers posed to open-source code publication by intellectual property rights, more research is possible in this area to find how intellectual property can accommodate IoT manufacturer cessation solutions.

Finally, the different possible solutions to IoT manufacturer cessation leads to potential for different regulatory schemes per IoT device. For instance, given the diverse impact of manufacturer cessation on IoT devices, a local-only alternative may fit certain devices (or impacts) better than mandated source code escrow, which brings certain financial requirements for the manufacturer. The Cyber Resilience Act already differs between levels of IoT devices based on their cybersecurity risk. A differentiation in pre-emptive obligation per type of IoT device, or perhaps based on price or other factors, could also be further explored.[98]

## 6) Conclusion

Internet of Things devices are omnipresent. These devices combine hardware and software and operate across diverse sectors. The EU legislators have recently regulated many aspects of Internet of Things products, primarily in relation to their overall cybersecurity and privacy practices.

---

98  See for instance the differing expectations surrounding privacy and security based on certain device characteristics in Lorenz Kustosch and others, 'Measuring Up to (Reasonable) Consumer Expectations: Providing an Empirical Basis for Holding IoT Manufacturers Legally Responsible', *Proceedings of the 32nd USENIX Security Symposium* (2023).

However, this paper shows that a legal gap remains: the position of the consumer after the IoT device's manufacturer ceases their operations.

The manufacturer's cessation leads to diverse problems, based on the reasons for the manufacturer to cease their operations and the type of products that they produce. A smart light system may still work manually after cessation, while other products rely strongly on mobile applications hosted by the developer which become unavailable. In general, this paper focused on manufacturers that have financial reasons for ceasing their operations (as this is most prevalent in practice) and products that ceased to functional entirely (as, in this case, a solution is most pressing to protect consumer interests).

Most EU legislation contains update obligations for modern products, to ensure that manufacturers do not simply disregard their devices when they want to focus their attention elsewhere. However, these update obligations, as exemplified by the Cyber Resilience Act and Sale of Goods Directive, rely on a key assumption: longevity of the manufacturer. When the manufacturer ceases to exist, their responsibility for updating or supporting their product disappears with them.

Other measures do show a hint of promise for consumers in the context of IoT manufacturer cessation, in the form of pre-emptive measures. The Cyber Resilience Act, Digital Operational Resilience Act and Data Governance Act include such pre-emptive measures, which require entities to concern themselves with the situation after they or third parties cease operations. This paper explores the potential of pre-emptive measures as a cornerstone for solving the IoT manufacturer cessation problem.

Pre-emptive measures are not sufficient by themselves. This paper adds two properties such measures should have: manufacturer-independency and collective control. Naturally, the consumer should no longer have to rely on the manufacturer for the solution, because the manufacturer ceases to exist. Furthermore, a system of collective control ensures that there is no longer a single point of failure: the consumer should not be stuck in a loop of failing organisations that ensure the continuity of their device.

Based on recent EU digital policy initiatives, the paper offers two examples of such manufacturer-independent collective control. First, it discusses interoperability, the capacity of the IoT device to exchange data with other devices and platforms. Interoperability is an important element of the General Data Protection Regulation and the recently enacted Data Act, Data Governance Act and Digital Markets Act. Based on a taxonomy of current interoperability research in the IoT domain, this article suggests that platform interoperability, in combination with a framework for updates, such as update pipeline interoperability or data exchange updates, could solve many of the issues brought by IoT manufacturer cessation. However, there are certainly restrictions on how much platform maintainers can offer in terms of updates.

Second, based on the Recital of the Cyber Resilience Act, the paper delves into open-source code publication, which allows developers other than the manufacturer to continue working on the device's code base. Source code publication currently faces several barriers, such as intellectual property rights, that first require further research or legislative work. However, given the interest in the Cyber Resilience Act, the article considers open-source code publication as a possible alternative to, or companion of, interoperability in the future.

The article closes with suggestions for future research, given the legislative and practical barriers identified for IoT manufacturer cessation, interoperability, and open-source code publication.